# Analytical expression for the harmonic Hall voltages in evaluating spin orbit torques

Masamitsu Hayashi

*National Institute for Materials Science, Tsukuba 305-0047, Japan*

**Solid understanding of current induced torques is key to the development of current and voltage controlled magnetization dynamics in ultrathin magnetic heterostructures. A versatile technique is needed to evaluate such torques in various systems. Here we examine the adiabatic (low frequency) harmonic Hall voltage measurement that has been recently developed to study current induced effective field that originate from the spin orbit effects. We analytically derive a form that can be used to evaluate the harmonic Hall voltages and extract relevant parameters in two representative systems, i.e. out of plane and in-plane magnetized systems. Contributions from the anomalous Hall and planar Hall effects are considered.**

*Email: hayashi.masamitsu@nims.go.jp

## I. Introduction

Application of current to systems with large spin orbit coupling in bulk or at interfaces may result in spin current generation via the spin Hall effect[1, 2] and/or current induced spin polarization (the Rashba-Edelstein effect).[3, 4] The generated spin current can act on nearby magnetic moments via spin transfer torque[5, 6] or exchange coupling[7, 8]. These effects are referred to as the "spin orbit torques",[8-14] which is to be distinguished from conventional spin transfer torque since the spin orbit coupling plays a critical role in generating the spin current.

Spin orbit torques are attracting great interest as they can lead to magnetization switching in geometries which were not possible with conventional spin transfer torque,[15, 16] and unprecedented fast domain wall motion[17, 18]. Solid understanding of how these torques arise is thus essential for developing devices utilizing spin orbit effects in ultrathin magnetic heterostructures.

Recently, it has been reported that an adiabatic (low frequency) harmonic Hall voltage measurement scheme, originally developed by Pi *et al.*,[19] can be used to evaluate the "effective magnetic field"[19-21] that generates the torque acting on the magnetic moments.[22-24] This technique has been used to evaluate the size and direction of the effective field in magnetic heterostructures. Using such technique, we have previously shown that the effective field shows a strong dependence on the layer thickness of Ta and CoFeB layers in Ta|CoFeB|MgO heterostructures.[22] The difference in the sign of the spin Hall angle between Ta and Pt has been probed and reported recently.[24] It has also been shown that there is a strong angular dependence (the angle between the magnetization and the current flow direction) of the effective field in Pt|Co|AlOx.[23] Non-local effects, i.e. spin current generated in a Pt layer can propagate through a Cu spacer and exert torques on the magnetic layer, have been probed using a similar technique.[25]

These results show that the adiabatic harmonic Hall voltage measurement is a powerful technique to study spin orbit torques in ultrathin magnetic heterostructures.

To provide a simple way to estimate the effective field using this important technique, here we derive an analytical form that describes the harmonic Hall voltages generated when a sinusoidal current is applied to the system. We assume that both the anomalous Hall and the planar Hall effects are present. In Ref. [23], Garello *et al.* have shown the importance of the planar Hall effect contribution to the harmonic Hall voltage. Here we derive a formula that is applicable to the expression which we have previously used[22] in evaluating the current induced effective field. In addition, we examine systems with out of plane and in-plane magnetization and compare the analytical solutions with numerical calculations based on a macrospin model.

**II. Analytical solutions**

  **A. Modulation amplitude of the magnetization angle**

The magnetic energy of the system can be expressed as

$$E = -K_{EFF} \cos^2 \theta + K_I \sin^2 \varphi \sin^2 \theta - \vec{M} \cdot \vec{H} . \tag{1}$$

where $K_{EFF}$ is the effective out of plane anisotropy energy and $K_I$ is the in-plane easy axis anisotropy energy. $\theta$ and $\varphi$ are the polar and azimuthal angles, respectively, of the magnetization (see Fig. 1 for the definition). $K_{EFF}$ and $K_I$ can be expressed as the following using the demagnetization coefficients $N_i$ ( $\sum_{i=X,Y,Z} N_i = 4\pi$ ) and the uniaxial magnetic anisotropy energy $K_U$:

$$K_{EFF} = K_U - \frac{1}{2}(N_Z - N_X)M_S^2$$
$$K_I = \frac{1}{2}(N_X - N_Y)M_S^2 \tag{2}$$

$K_U$ is defined positive for out of plane magnetic easy axis. The direction of the magnetization $\vec{M}$ and the external magnetic field $\vec{H}$ are expressed using $\theta$ and $\varphi$ as:

$$\vec{M} = M_s \hat{m}, \quad \hat{m} = (\sin\theta\cos\varphi, \sin\theta\sin\varphi, \cos\theta), \tag{3}$$

$$\vec{H} = H(\sin\theta_H \cos\varphi_H, \sin\theta_H \sin\varphi_H, \cos\theta_H). \tag{4}$$

$M_s$ is the saturation magnetization, $\hat{m}$ is a unit vector representing the magnetization direction and $H$ represents the magnitude of the external magnetic field.

The equilibrium magnetization direction ($\theta_0$, $\varphi_0$) is calculated using the following two equations:

$$\frac{\partial E}{\partial \theta} = -\left(-K_{EFF} + K_I \sin^2\varphi_0\right)\sin 2\theta_0 - M_S H\left(\cos\theta_0 \sin\theta_H (\cos\varphi_0 \cos\varphi_H + \sin\varphi_0 \sin\varphi_H) - \sin\theta_0 \cos\theta_H\right) = 0, \tag{5a}$$

$$\frac{\partial E}{\partial \varphi} = -K_I \sin^2\theta_0 \sin 2\varphi_0 - M_S H \sin\theta_0 \sin\theta_H \sin(\varphi_H - \varphi_0) = 0. \tag{5b}$$

Equations (5a) and (5b) can be solved to obtain ($\theta_0$, $\varphi_0$), which will be discussed later. To simplify notations, we define $H_K \equiv \dfrac{2K_{EFF}}{M_S}$ and $H_A \equiv \dfrac{2K_I}{M_S}$.

When current is passed to the device under test, current induced effective field $\Delta H_{X,Y,Z}$, including the Oersted field, can modify the magnetization angle from its equilibrium value ($\theta_0$, $\varphi_0$). The change in the angle, termed the modulation amplitudes $(\Delta\theta, \Delta\varphi)$ hereafter, can be calculated using the following equations:[23, 26]

$$\Delta\theta = \frac{\delta\theta}{\delta H_X}\Delta H_X + \frac{\delta\theta}{\delta H_Y}\Delta H_Y + \frac{\delta\theta}{\delta H_Z}\Delta H_Z, \tag{8a}$$

$$\Delta\varphi = \frac{\delta\varphi}{\delta H_X}\Delta H_X + \frac{\delta\varphi}{\delta H_Y}\Delta H_Y + \frac{\delta\varphi}{\delta H_Z}\Delta H_Z. \tag{8b}$$

Here $\dfrac{\delta\theta}{\delta H_i}$ and $\dfrac{\delta\varphi}{\delta H_i}$ ($i$=X, Y, Z) represent the degree of change in the angles when a field is applied along one of the (Cartesian) axes. To calculate $\dfrac{\delta\theta}{\delta H_i}$ and $\dfrac{\delta\varphi}{\delta H_i}$, we use the following relations derived from Eqs. (5a) and (5b):

$$\dfrac{\delta}{\delta H_i}\left(\dfrac{\partial E}{\partial \theta}\right) = 0 = \left[\left(K_{EFF} - K_I \sin^2\varphi_0\right)2\cos 2\theta_0 - M_S\left(-H_X \sin\theta_0 \cos\varphi_0 - H_Y \sin\theta_0 \sin\varphi_0 - H_Z \cos\theta_0\right)\right]\dfrac{\delta\theta}{\delta H_i} \quad (9a)$$
$$+ \left[-2K_I \sin\varphi_0 \cos\varphi_0 \sin 2\theta_0 - M_S\left(-H_X \cos\theta_0 \sin\varphi_0 + H_Y \cos\theta_0 \cos\varphi_0\right)\right]\dfrac{\delta\varphi}{\delta H_i} - M_S f_i$$

$$\dfrac{\delta}{\delta H_i}\left(\dfrac{\partial E}{\partial \varphi}\right) = 0 = \left[-K_I \cos\theta_0 \sin 2\varphi_0\right]\dfrac{\delta\theta}{\delta H_i} \quad (9b)$$
$$+ \left[-2K_I \sin\theta_0 \cos 2\varphi_0 + M_S\left(H_X \cos\varphi_0 + H_Y \sin\varphi_0\right)\right]\dfrac{\delta\varphi}{\delta H_i} + M_S g_i$$

where

$$f_i = \begin{bmatrix} \cos\theta_0 \cos\varphi_0 \\ \cos\theta_0 \sin\varphi_0 \\ -\sin\theta_0 \end{bmatrix}, \quad g_i = \begin{bmatrix} \sin\varphi_0 \\ -\cos\varphi_0 \\ 0 \end{bmatrix}.$$

The coupled equations (9a) and (9b) can be solved for $\dfrac{\delta\theta}{\delta H_i}$ and $\dfrac{\delta\varphi}{\delta H_i}$, which reads:

$$\dfrac{\delta\theta}{\delta H_i} = \dfrac{1}{F_1}\left[f_i - C g_i\right] \quad (10a)$$

$$\dfrac{\delta\varphi}{\delta H_i} = \dfrac{1}{F_1 F_2}\left[\dfrac{1}{2} f_i H_A \cos\theta_0 \sin 2\varphi_0 - g_i\left[\left(H_K - H_A \sin^2\varphi_0\right)\cos 2\theta_0 + \vec{H}\cdot\hat{m}\right]\right] \quad (10b)$$

$$F_1 \equiv \left(H_K - H_A \sin^2\varphi_0\right)\cos 2\theta_0 + \vec{H}\cdot\hat{m} - \dfrac{1}{2}C H_A \cos\theta_0 \sin 2\varphi_0$$

$$F_2 \equiv -H_A \sin\theta_0 \cos 2\varphi_0 + H_X \cos\varphi_0 + H_Y \sin\varphi_0$$

$$C \equiv \dfrac{1}{F_2}\left[\dfrac{1}{2} H_A \sin 2\theta_0 \sin 2\varphi_0 + \left(-H_X \sin\varphi_0 + H_Y \cos\varphi_0\right)\cos\theta_0\right]$$

Substituting Eqs. (10a) and (10b) into Eqs. (8a) and (8b) gives the following expressions for the modulation amplitudes of the magnetization angle:

$$\Delta\theta = \frac{1}{F_1}\left[\left(\cos\theta_0\cos\varphi_0 - C\sin\varphi_0\right)\Delta H_X + \left(\cos\theta_0\sin\varphi_0 + C\cos\varphi_0\right)\Delta H_Y - \left(\sin\theta_0\right)\Delta H_Z\right] \quad (11a)$$

$$\Delta\varphi = \frac{1}{F_1 F_2}\left[H_A\left(\cos^2\theta_0\cos^2\varphi_0 + \cos 2\theta_0 \sin^2\varphi_0\right) - H_K \cos 2\theta_0 - \vec{H}\cdot\hat{m}\right]\Delta H_X \sin\varphi_0$$
$$+ \frac{1}{F_1 F_2}\left[H_A \sin^2\theta_0 \sin^2\varphi_0 + H_K \cos 2\theta_0 + \vec{H}\cdot\hat{m}\right]\Delta H_Y \cos\varphi_0 \quad (11b)$$
$$+ \frac{1}{F_1 F_2}\left[-\frac{1}{2}H_A \sin\theta_0 \cos\theta_0 \sin 2\varphi_0\right]\Delta H_Z$$

Equations (11a) and (11b) are valid for any equilibrium magnezation direction and are general for arbitrary values of each parameter (no approximation made).

### B. Expression for the Hall voltage

The Hall voltage typically contains contributions from the anomalous Hall effect (AHE) and the planar Hall effect (PHE). We define $\Delta R_A$ and $\Delta R_P$ as the change in the Hall resistance due to the AHE and PHE, respectively, when the magnetization direction reverses. Assuming a current flow along the x-axis, the Hall resistance $R_{XY}$ is expressed as:

$$R_{XY} = \frac{1}{2}\Delta R_A \cos\theta + \frac{1}{2}\Delta R_P \sin^2\theta \sin 2\varphi \quad (12)$$

If we substitute $\theta=\theta_0+\Delta\theta$, $\varphi=\varphi_0+\Delta\varphi$ and assume $\Delta\theta<<1$ and $\Delta\varphi<<1$, Eq. (12) can be expanded to read

$$R_{XY} \approx \frac{1}{2}\Delta R_A \left(\cos\theta_0 - \Delta\theta\sin\theta_0\right) + \frac{1}{2}\Delta R_P \left(\sin^2\theta_0 + \Delta\theta\sin 2\theta_0\right)\left(\sin 2\varphi_0 + 2\Delta\varphi\cos 2\varphi_0\right) \quad (13)$$

The Hall voltage $V_{XY}$ is a product of the Hall resistance $R_{XY}$ and the current $I$ passed along the device, i.e.

$$V_{XY} = R_{XY} I \quad (14)$$

When a sinusoidal current (excitation amplitude $\Delta I$, frequency $\omega$) is applied, the current induced effective field oscillates in sync with the current. Thus $\Delta H_{X,Y,Z}$ in Eqs. (11a) and (11b) needs to be replaced with $\Delta H_{X,Y,Z} \sin \omega t$. Substituting Eq. (13) into Eq. (14) gives:

$$
\begin{aligned}
V_{XY} &= V_0 + V_\omega \sin \omega t + V_{2\omega} \cos 2\omega t \\
V_0 &= \frac{1}{2}(B_\theta + B_\varphi)\Delta I, \\
V_\omega &= A\Delta I, \\
V_{2\omega} &= -\frac{1}{2}(B_\theta + B_\varphi)\Delta I \\
A &= \frac{1}{2}\Delta R_A \cos\theta_0 + \frac{1}{2}\Delta R_P \sin^2\theta_0 \sin 2\varphi_0 \\
B_\theta &= \frac{1}{2}(-\Delta R_A \sin\theta_0 + \Delta R_P \sin 2\theta_0 \sin 2\varphi_0)\Delta\theta \\
B_\varphi &= \Delta R_P \sin^2\theta_0 \cos 2\varphi_0 \Delta\varphi
\end{aligned}
\quad (15)
$$

As evident in Eq. (15), the second harmonic Hall voltage $V_{2\omega}$ contains information of $\Delta H_{X,Y,Z}$ through $\Delta\theta$ and $\Delta\varphi$. Note that Eq. (15) describes the harmonic Hall voltage in the limit of small $\Delta\theta$ and $\Delta\varphi$.

**C. Relation between the current induced effective field and spin torque**

To illustrate the relationship between the current induced effective field $\Delta H_{X,Y,Z}$ and the conventional spin torque terms, $\Delta H_{X,Y,Z}$ can be added, as a vector $\Delta \vec{H}$, in the Landau-Lifshitz-Gilbert (LLG) equation:

$$\frac{\partial \hat{m}}{\partial t} = -\gamma \hat{m} \times \left( \frac{\partial E}{\partial \vec{M}} + \Delta \vec{H} \right) + \alpha \hat{m} \times \frac{\partial \hat{m}}{\partial t} \quad (16a)$$

Here α is the Gilbert damping constant, γ is the gyromagnetic ratio, $\frac{\partial E}{\partial \vec{M}}$ is the effective magnetic field that includes external, exchange, anisotropy and demagnetization fields.

Equation (16a) can be compared to the LLG equation with the two spin torque terms.

$$\frac{\partial \hat{m}}{\partial t} = -\gamma \hat{m} \times \left( \frac{\partial E}{\partial \vec{M}} + a_J (\hat{m} \times \hat{p}) + b_J \hat{p} \right) + \alpha \hat{m} \times \frac{\partial \hat{m}}{\partial t} \quad (16b)$$

Here $\hat{p}$ represents the magnetization direction of the "reference layer" in spin valve nanopillars/magnetic tunnel junctions, $a_J$ and $b_J$ correspond to the Slonczweski-Berger[5, 6] (STT) and the field-like effective fields[27], respectively. Comparing Eqs. (16a) and (16b), we can decompose the current induced effective field $\Delta \vec{H}$ into two terms, $\Delta \vec{H} \equiv \Delta \vec{H}_{STT} + \Delta \vec{H}_{FL}$, where $\Delta \vec{H}_{STT} \equiv a_J \hat{m} \times \hat{p}$ and $\Delta \vec{H}_{FL} \equiv b_J \hat{p}$. The STT-term $\Delta \vec{H}_{STT}$ depends on the magnetization direction whereas the field like term $\Delta \vec{H}_{FL}$ is independent of $\hat{m}$. Note that the harmonic Hall voltage measures $\Delta \vec{H}$ and not the torque ($-\gamma (\hat{m} \times \Delta \vec{H})$); thus one can identify whether the effective field is STT-like or field-like by measuring its dependence on the magnetization direction.

For the numerical calculations, we use $\hat{p} = (0,1,0)$ as this represents the spin direction of the electrons entering the CoFeB layer via the spin Hall effect in Ta when current is passed along the +x axis for Ta|CoFeB|MgO heterostructures. In the following, we consider two representative cases, systems with out of plane and in-plane magnetization.

**III. Approximate expressions for the harmonic Hall voltages**

## A. Out of plane magnetization systems

We first consider a system where the magnetization points along the film normal owing to its perpendicular magnetic anisotropy. To obtain analytical solutions for the harmonic Hall voltages, we make several approximations. First, to solve Eqs. (5a) and (5b) analytically, we assume that the in-plane uniaxial anisotropy is small compared to the external field, i.e. $|H_A| \ll |H|$. Equation (5b) then gives $\varphi_0 = \varphi_H$. Next we assume that the equilibrium magnetization direction does not deviate much from the z-axis, i.e. $\theta_0 = \theta_0'$ $(\theta_0' \ll 1)$ for $\vec{M}$ along $+\hat{z}$ and $\theta_0 = \pi - \theta_0'$ $(\theta_0' \ll 1)$ for $\vec{M}$ along $-\hat{z}$. Keeping terms that are linear with $\theta_0'$, Eqs. (5a) and (5b) give:

$$\theta_0' = \pm \frac{H \sin\theta_H}{\pm(H_K - H_A \sin^2\varphi_H) + H\cos\theta_H}, \quad \varphi_0 = \varphi_H \tag{17}$$

The $\pm$ sign corresponds to the case for $\vec{M}$ pointing along $\pm\hat{z}$. Assuming $\theta_0' \ll 1$ and $|H_A| \ll |H|, |H_K|$, expressions (11a) and (11b) can be simplified to read:

$$\Delta\theta \approx \frac{1}{H_K \pm H\cos\theta_H}\left[\pm\Delta H_{IN} + \theta_0'\left(-\Delta H_Z \pm \frac{H\sin\theta_H}{H_K \pm H\cos\theta_H}\Delta H_{IN} \pm \frac{H_A \sin 2\varphi_H}{H\sin\theta_H}\Delta\tilde{H}_{IN}\right)\right] \tag{18a}$$

$$\Delta\varphi \approx \frac{\Delta\tilde{H}_{IN}}{H\sin\theta_H} \mp \frac{\frac{1}{2}H_A \sin\theta_0' \cos\theta_0' \sin 2\varphi_H \Delta H_Z}{\left[H_K \cos 2\theta_0' + H(\pm\cos\theta_H \cos\theta_0' + \sin\theta_H \sin\theta_0')\right]\left[-H_A \sin\theta_0' \cos 2\varphi_H + H\sin\theta_H\right]}$$
(18b)

where we have defined the following fields:

$$\begin{aligned}\Delta H_{IN} &\equiv \Delta H_X \cos\varphi_H + \Delta H_Y \sin\varphi_H \\ \Delta\tilde{H}_{IN} &\equiv -\Delta H_X \sin\varphi_H + \Delta H_Y \cos\varphi_H\end{aligned} \tag{19}$$

Substituting Eqs. (17), (18a) and (18b) into Eq. (15) gives:

$$V_\omega \approx \frac{1}{2}\left[\pm \Delta R_A + \left(\mp\frac{1}{2}\Delta R_A + \Delta R_P \sin 2\varphi_H\right)\left(\frac{H\sin\theta_H}{\pm H_K + H\cos\theta_H}\right)^2\right]\Delta I \qquad (20a)$$

$$V_{2\omega} \approx -\frac{1}{4}\left[\left[\left(\mp\Delta R_A + 2\Delta R_P \sin 2\varphi_H\right)\Delta H_{IN} + 2\Delta R_P \cos 2\varphi_H \Delta \tilde{H}_{IN}\right]\frac{H\sin\theta_H}{\left(H_K \pm H\cos\theta_H\right)^2}\right]\Delta I \qquad (20b)$$

We have neglected higher order terms with $\theta'_0$. In addition, if we consider cases where the external field is directed along one of the Cartesian coordinate axes (along x- or y-axis), terms with $\sin(2\varphi_H)$ will also vanish. Then Eqs. (20a) and (20b) can be simplified to read:

$$V_\omega \approx \pm\frac{1}{2}\Delta R_A\left[1 - \frac{1}{2}\left(\frac{H\sin\theta_H}{\pm H_K + H\cos\theta_H}\right)^2\right]\Delta I \qquad (21a)$$

$$V_{2\omega} \approx -\frac{1}{4}\left[\left(\mp\Delta R_A \Delta H_{IN} + 2\Delta R_P \cos 2\varphi_H \Delta \tilde{H}_{IN}\right)\frac{H\sin\theta_H}{\left(H_K \pm H\cos\theta_H\right)^2}\right]\Delta I$$

$$= -\frac{1}{4}\left[\left[\left(\mp\Delta R_A \Delta H_X + 2\Delta R_P \cos 2\varphi_H \Delta H_Y\right)\cos\varphi_H + \left(\mp\Delta R_A \Delta H_Y - 2\Delta R_P \cos 2\varphi_H \Delta H_X\right)\sin\varphi_H\right]\frac{H\sin\theta_H}{\left(H_K \pm H\cos\theta_H\right)^2}\right]\Delta I$$
(21b)

Equation (21b) shows that the planar Hall effect mixes the signal from different components of the current induced effective field.[23] For systems with negligible PHE, $\Delta H_X$ ($\Delta H_Y$) can be determined by measuring $V_{2\omega}$ as a function of the external in-plane field directed along the x- (y-) axis. However, if the PHE becomes comparable to the size of AHE, contribution from the orthogonal component appears in $V_{2\omega}$ via the PHE. When $\Delta R_P$ is larger than half of $\Delta R_A$, $\Delta H_Y$ ($\Delta H_X$) becomes the dominant term in $V_{2\omega}$ for field sweep along the x- (y-) axis. Thus to estimate the effective field components accurately in systems with non-negligble PHE, one needs to measure $V_{2\omega}$ in two orthogonal directions and analytically calculate each component, as described below.

We follow the procedure used previously[22] to eliminate the prefactors that are functions of $\Delta I$ and $H_K$ in Eqs. (21a) and (21b). $\theta_H=\pi/2$ is substituted in Eqs. (21a) and (21b) since the external field is swept along the film plane. The respective curvature and slope of $V_\omega$ and $V_{2\omega}$ versus the external field are calculated to obtain their ratio $B$:

$$B \equiv \left( \frac{\partial V_{2\omega}}{\partial H} \bigg/ \frac{\partial^2 V_\omega}{\partial H^2} \right) = -\frac{1}{2}\left[ \left( \Delta H_X \mp 2\frac{\Delta R_P}{\Delta R_A}\cos 2\varphi_H \Delta H_Y \right)\cos \varphi_H + \left( \Delta H_Y \pm 2\frac{\Delta R_P}{\Delta R_A}\cos 2\varphi_H \Delta H_X \right)\sin \varphi_H \right] \quad (22)$$

We define $B_X \equiv \left( \frac{\partial V_{2\omega}}{\partial H} \bigg/ \frac{\partial^2 V_\omega}{\partial H^2} \right)\bigg|_{\varphi_H=0}$ and $B_Y \equiv \left( \frac{\partial V_{2\omega}}{\partial H} \bigg/ \frac{\partial^2 V_\omega}{\partial H^2} \right)\bigg|_{\varphi_H=\frac{\pi}{2}}$, which correspond to $B$ measured for $\varphi_H=0$ and $\varphi_H=\pi/2$, respectively, and $\xi \equiv \frac{\Delta R_P}{\Delta R_A}$, which is the ratio of the PHE and AHE resistances. Finally, we obtain:

$$\begin{aligned}\Delta H_X &= -2\frac{(B_X \pm 2\xi B_Y)}{1-4\xi^2} \\ \Delta H_Y &= -2\frac{(B_Y \pm 2\xi B_X)}{1-4\xi^2}\end{aligned} \quad (23)$$

Equation (23) provides a simple method to obtain the effective field under circumstances where both AHE and PHE contribute to the Hall signal. When PHE is negligible, $\xi=0$ and we recover the form derived previously.[22]

**B. In-plane magnetization systems**

We next consider systems with in-plane magnetization (easy axis is along the x-axis). We make the same approximation used in the previous section, $|H_A|<<|H|$, $|H_K|$ (note that $H_K<0$ for in-plane magnetized systems). This may not apply to systems where the shape anisotropy is large, e.g. in high aspect ratio elements such as nanowires. However, lifting this restriction will introduce more complexity in the analysis of the harmonic voltage, in particular, because one cannot make the assumption $\varphi_0 \sim \varphi_H$. Here we limit systems with small $H_A$.

To obtain each component of the effective field (the STT and the field-like terms), the external field must be swept along two directions orthogonal to the magnetization direction. From symmetry arguments, the out of plane field sweep gives information of the field-like term whereas the in-plane transverse field sweep (transverse to the magnetization direction) provides the STT term. However, the assumption $|H_A|<<|H|$ used here causes the magnetization to rotate along the transverse field direction as soon as $H$ is larger than $H_A$, thus hindering evaluation of the STT term. To circumvent this difficulty, we make use of the anisotropic magnetoresistance (AMR) of the magnetic material and measure the longitudinal voltage ($V_{XX}$), which in turn provides information of the STT term. We start from the Hall voltage measurements and move on to the longitudinal resistance measurement in the next section.

### 1. Harmonic Hall voltage and the field-like term

We assume that the equilibrium magnetization direction does not deviate much from the film plane, i.e. $\theta_0 = \frac{\pi}{2} + \theta'_0$ ($\theta'_0 \ll 1$) for $\vec{M}$ pointing along one direction within the film plane and $\theta_0 = \frac{\pi}{2} - \theta'_0$ ($\theta'_0 \ll 1$) for the case where the direction of $\vec{M}$ is reversed from the former. Note that $\varphi_0 = \varphi_H$ for $|H_A| \ll |H|$ (see Eq. (5b)). Keeping terms that are linear with $\theta'_0$, Eqs. (5a) and

(5b) give:

$$\theta_0' = \pm \frac{H \cos\theta_H}{(H_K - H_A \sin^2\varphi_H) - H \sin\theta_H}, \quad \varphi_0 = \varphi_H. \tag{24}$$

The ± sign corresponds to the case for $\theta_0 = \theta_0' \pm \frac{\pi}{2}$. Assuming $\theta_0' \ll 1$ and $|H_A| \ll |H|, |H_K|$, expressions (11a) and (11b) can be simplified to read:

$$\Delta\theta \approx \frac{1}{-H_K + H\sin\theta_H}\left[-\Delta H_Z \mp \theta_0\left(\Delta H_{IN} + \frac{H_A \sin 2\varphi_H}{H\sin\theta_H}\Delta\tilde{H}_{IN} + \frac{H\cos\theta_H}{-H_K + H\sin\theta_H}\Delta H_Z\right)\right] \tag{25a}$$

$$\Delta\varphi \approx \frac{\Delta\tilde{H}_{IN}}{-H_A\cos\theta_0'\cos 2\varphi_H + H\sin\theta_H}$$

$$\mp \frac{\frac{1}{2}H_A\cos\theta_0'\sin\theta_0'\sin 2\varphi_H \Delta H_Z}{\left[-H_K\cos 2\theta_0' + H(\sin\theta_H\cos\theta_0' \mp \cos\theta_H\sin\theta_0')\right]\left[-H_A\cos\theta_0'\cos 2\varphi_H + H\sin\theta_H\right]} \tag{25b}$$

Substituting Eqs. (24), (25a) and (25b) into Eq. (15) gives:

$$V_\omega \approx \frac{1}{2}\left[-\Delta R_A \frac{H\cos\theta_H}{H_K - H\sin\theta_H} + \Delta R_P \sin 2\varphi_H\right]\Delta I \tag{26}$$

$$V_{2\omega} \approx -\frac{1}{4}\left[\frac{\Delta R_A}{H_K - H\sin\theta_H}\left(-\Delta H_Z + \Delta H_{IN}\frac{H\cos\theta_H}{H_K - H\sin\theta_H}\right) + 2\Delta R_P \cos 2\varphi_H\left(\Delta\tilde{H}_{IN}\frac{1}{-H_A\cos 2\varphi_H + H\sin\theta_H}\right)\right]\Delta I \tag{27}$$

We have neglected higher order terms with $\theta_0'$. The second term (the $\Delta R_P$ term) in Eq. (27) dominates the second harmonic signal for samples with large $|H_K|$. Thus for a typical in-plane magnetized samples, one can ignore the first term in Eq. (27) to obtain:

$$V_{2\omega} \approx -\frac{1}{2}\left[\Delta R_P \cos 2\varphi_H\left(\frac{-\Delta H_X \sin\varphi_H + \Delta H_Y \cos\varphi_H}{-H_A \cos 2\varphi_H + H\sin\theta_H}\right)\right]\Delta I \tag{28}$$

It should be noted that the sign of $V_{2\omega}$ at a given $H$, which determines the sign of the effective field, changes when the field is slightly tilted in one direction or the other (e.g. $\theta_H \sim 1$ or $-1$ deg). When the field is directed exactly along the film normal ($\theta_H=0$), the field dependence of $V_{2\omega}$ vanishes.

For in-plane magnetized samples, the STT *effective field* is directed along the film normal (not to be confused with the STT *torque* that points along the film plane). That is, $\Delta H_Z$ corresponds to the STT term. See the inset of Fig. 3(b) in which we show the three effective field components $\Delta H_{X,Y,Z}$ when the magnetization and the current flow directions are along +x and the incoming electrons' spin polarization is set to +y ($\hat{p}=(0,1,0)$). However, in Eq. (28), $\Delta H_Z$ does not show up in the expression, illustrating the difficulty of evaluating the STT term in this geometry. To extract the STT term, one needs to measure the harmonic longitudinal voltages, which are shown in the next section. Here we proceed to obtain the field-like term $\Delta H_Y$.

Using Eqs. (26) and (28), we derive a simple formula, similar to that shown in Eq. (23), to estimate $\Delta H_Y$. If the external field is slightly tilted from the film normal, it is preferable to apply the in-plane field component along the magnetic easy axis ($\varphi_H=0$ deg) to unambiguously set the equilibrium magnetization azimuthal angle $\varphi_0$. Substituting $\varphi_H=0$ in Eqs. (26a) and (28) gives:

$$\frac{\partial V_\omega}{\partial H} \approx -\frac{1}{2}\Delta R_A \Delta I \frac{\cos\theta_H}{H_K} \tag{29a}$$

$$\frac{\partial(1/V_{2\omega})}{\partial H} \approx -\frac{2\sin\theta_H}{\Delta R_P \Delta I \Delta H_Y} \tag{29b}$$

Thus $\Delta H_Y$ (the field-like term) can be obtained by:

$$\Delta H_Y \approx \frac{\sin 2\theta_H}{2\xi H_K}\left[1\bigg/\left(\frac{\partial V_\omega}{\partial H}\right)\left(\frac{\partial(1/V_{2\omega})}{\partial H}\right)\right] \tag{30}$$

Unlike the case derived in the previous section (Eq. (23)) here one needs to substitute $H_K$ and $\theta_H$ to calculate the effective field. This is because the field dependence of the first harmonic voltage is primarily determined by the change in the magnetization direction along the z-axis (relevant anisotropy is $H_K$) whereas that of the second harmonic voltage is dominated by the magnetization angular change within the film plane (relevant anisotropy is $H_A$); therefore, taking the ratio of the two will not cancel out $H_K$ (and $\theta_H$). Since $\Delta H_Y$ scales with the tilt angle $\theta_H$, one needs to determine the value of $\theta_H$ in good precision.

### 2. Harmonic longitudinal voltage and the STT term

As noted above, in order to obtain the STT term, one can make use of the AMR effect, if any, of the magnetic material. The longitudinal resistance $R_{XX}$ is expressed as:

$$R_{XX} = R_0 + \frac{1}{2}\Delta R_{MR} \sin^2\theta \cos^2\varphi \tag{31}$$

where $R_0$ and $\Delta R_{MR}$ are, respectively, the resistance independent of the magnetization direction and the change in the resistance due to the AMR effect. Current is assumed to flow along the x-axis. We substitute $\theta=\theta_0+\Delta\theta$, $\varphi=\varphi_0+\Delta\varphi$ into Eq. (31) and assume $\Delta\theta\ll1$ and $\Delta\varphi\ll1$, which then reads:

$$R_{XX} \approx R_0 + \frac{1}{2}\Delta R_{MR}\left(\sin^2\theta_0\cos^2\varphi_0 + \Delta\theta\sin 2\theta_0\cos^2\varphi_0 - \Delta\varphi\sin^2\theta_0\sin 2\varphi_0 - (\Delta\theta^2+\Delta\varphi^2)\sin^2\theta_0\cos^2\varphi_0\right)$$
(32)

Here, we have kept the 2nd order terms that scales with $\Delta\theta^2$ and $\Delta\varphi^2$ to show that these terms cannot be neglected when the current induced effective field $\Delta H_{X,Y,Z}$ (that determines the magnitude of $\Delta\theta$ and $\Delta\varphi$) becomes larger. This is because, for a typical geometry that would be employed here (current flow and the equilibrium magnetization directions pointing along +x),

sin($2\theta_0$)~0 and sin($2\varphi_0$)~0, and thus the linear terms (that scales with $\Delta\theta$ and $\Delta\varphi$) can be smaller than the 2nd order terms when $\Delta\theta$ or $\Delta\varphi$ is larger than a critical value. As a consequence, there is a limit in $\Delta H_{X,Y,Z}$ above which we cannot neglect the 2nd order terms and this limit is much smaller than the other geometries described previously. For simplicity, we only consider the small limit of $\Delta H_{X,Y,Z}$ here.

Again, for in-plane magnetized systems, we substitute $\theta_0 = \frac{\pi}{2} + \theta_0'$ ($\theta_0' \ll 1$) for $\vec{M}$ pointing along one direction within the film plane and $\theta_0 = \frac{\pi}{2} - \theta_0'$ ($\theta_0' \ll 1$) when $\vec{M}$ is reversed. With $|H_A|<<|H|$, $\varphi \sim \varphi_H$, Eq. (32) reads:

$$R_{XX} \approx R_0 + \frac{1}{2}\Delta R_{MR}\left(\cos^2\theta_0' \cos^2\varphi_H \mp \Delta\theta \sin 2\theta_0' \cos^2\varphi_H - \Delta\varphi \cos^2\theta_0 \sin 2\varphi_H \right) \tag{33}$$

Substituting Eqs. (25a) and (25b) into Eq. (33) results in the following expression:

$$R_{XX} \approx R_0 + \frac{1}{2}\Delta R_{MR}\left(\cos^2\theta_0' \cos^2\varphi_H \mp \sin 2\theta_0' \cos^2\varphi_H \frac{\pm\Delta H_{IN}\sin\theta_0' + \Delta H_Z \cos\theta_0'}{H_K \cos 2\theta_0' - H\sin(\theta_H \mp \theta_0')}\right) \tag{34}$$

We have dropped the terms with $\sin 2\varphi_H$, which is zero when $\varphi_H=0$ or $\pi$. Substitution of Eq. (24) into Eq. (34) leads to:

$$R_{XX} \approx R_0 + \frac{1}{2}\Delta R_{MR}\cos^2\varphi_H \left[1 - \left(\frac{H\cos\theta_H}{-H_K + H\sin\theta_H}\right)^2 + \frac{2H\cos\theta_H}{-H_K + H\sin\theta_H}\frac{\Delta H_Z}{H_K}\right] \tag{35}$$

where we have omitted terms that are small. A sinusoidal excitation current $I = \Delta I \sin\omega t$ is applied to the device and the resulting voltage is expressed as:

$$V_{XX} = R_{XX}I = R_{XX}\Delta I \sin\omega t \tag{36}$$

Again, assuming that the current induced effective fields are in sync with the excitation current, Eqs. (35) and (36) give:

$$V_{XX} = V_0^{XX} + V_\omega^{XX} \sin \omega t + V_{2\omega}^{XX} \cos 2\omega t$$

$$V_0^{XX} = \frac{1}{2} B \Delta I$$

$$V_\omega^{XX} = A \Delta I$$

$$V_{2\omega}^{XX} = -\frac{1}{2} B \Delta I \tag{37}$$

$$A = R_0 + \frac{1}{2} \Delta R_{MR} \cos^2 \varphi_H \left[ 1 - \left( \frac{H \cos \theta_H}{-H_K + H \sin \theta_H} \right)^2 \right]$$

$$B = \frac{1}{2} \Delta R_{MR} \cos^2 \varphi_H \frac{2H \cos \theta_H}{-H_K + H \sin \theta_H} \frac{\Delta H_Z}{H_K}$$

These expressions are similar to those of Eqs. (21a) and (21b). The ratio of the field derivatives of the first and second harmonic signals directly provides the current induced effective field (STT term):

$$\Delta H_Z = -2 \left( \frac{\partial V_{2\omega}^{XX}}{\partial H} \bigg/ \frac{\partial^2 V_\omega^{XX}}{\partial H^2} \right) \tag{38}$$

Unlike the harmonic Hall voltage measurements, where both the AHE and the PHE contribute to the signal, here the longitudinal voltage $V_{XX}$ is determined solely by the AMR effect. Equations (30) and (38) show that combination of the Hall and longitudinal voltage measurements can provide means to evaluate both components of the effective field for in-plane magnetized samples.

## IV. Comparison to numerical calculations

The analytical solutions derived above are compared to numerical calculations. We solve Eq. (16b) numerically to obtain the equilibrium magnetization direction and the associated harmonic voltage signals. A macrospin model[28, 29] is used to describe the system. Substituting Eq. (1) into Eq. (16b), the following differential equations are obtained:

$$\frac{1+\alpha^2}{\gamma}\frac{\partial \theta}{\partial t} = \alpha h_\theta + h_\varphi$$

$$\frac{1+\alpha^2}{\gamma}\frac{\partial \varphi}{\partial t}\sin\theta = -h_\theta + \alpha h_\varphi$$

$$h_\theta = a_J \cos\varphi + b_J \cos\theta \sin\varphi + h_X \cos\theta \cos\varphi + h_Y \cos\theta \sin\varphi - h_Z \sin\theta$$

$$h_\varphi = a_J \cos\theta \sin\varphi + b_J \cos\varphi - h_X \sin\varphi + h_Y \cos\varphi \qquad (46)$$

$$h_X = H_X - M_S N_X \sin\theta \cos\varphi$$

$$h_Y = H_Y - M_S N_Y \sin\theta \sin\varphi$$

$$h_Z = H_Z - (M_S N_Z - H_K)\cos\theta$$

The two coupled differential equations are numerically solved to obtain the equilibrium magnetization direction when both the external and the current induced effective fields are turned on. To mimic the experimental setup, a sinusoidal current is passed along the x-axis and the resulting Hall and longitudinal voltages are evaluated. Contributions from the anomalous Hall effect (AHE) and the planar Hall effect (PHE) are considered for the Hall voltage and that from the anisotropic magnetoresistance (AMR) is taken into account for the longitudinal voltage. For a given time during one cycle of the sinusoidal current application, we calculate the equilibrium magnetization direction and the corresponding Hall and longitudinal voltages. One cycle is divided into two hundred time steps to obtain the temporal variation of the Hall and longitudinal voltages. The calculated voltages are fitted with Eq. (15): i.e. $V_{XY} = V_0 + V_\omega \sin\omega t + V_{2\omega} \cos 2\omega t$, to obtain the first and second harmonic signals. We compare the numerical results with the analytical solutions derived in the previous sections.

### A. Out of plane magnetization systems

Figure 2 shows results for the out of plane magnetized samples. The equilibrium magnetization angle ($\theta_0$) with respect to the film normal (a), first (b) and second (d) harmonic Hall voltages are plotted against an in-plane field directed along the current flow direction (i.e. along the x-axis).

The transverse field (directed along the y-axis) dependence of the second harmonic Hall voltage is shown in Fig. 2(f); the corresponding magnetization angle ($\theta_0$) and the first harmonic Hall voltage are the same with those shown in (a) and (b), respectively. The parameters used here mimic the system of Ta|CoFeB|MgO heterostructures[22] (see Fig. 2 caption for the details). The open symbols represent results from the numerical calculations (squares: magnetization pointing +z, circles: magnetization along -z) whereas the solid/dashed lines correspond to the analytical results. As evident, the analytical solutions agree well with the numerical results.

The x, y, z components of the current induced effective field are shown in Fig. 2(c) and 2(e) when the in-plane field is swept along x- and y-axis, respectively. We use $\hat{p}=(0,1,0)$, $a_J$=3 Oe and $b_J$=3 Oe. The difference of $\Delta H_{X,Y,Z}$ in (c) and (e) is due to the difference in the magnetization azimuthal angle: $\varphi_0$~0 for (c) and $\varphi_0$~$\pi$/2 for (e). As shown in Fig. 2(c,e), at low magnetic field, one can consider $\Delta H_X$ and $\Delta H_Y$ as $a_J$ and $b_J$, respectively. We test the validity of Eq. (23) by fitting the external field dependence of the first and second harmonic voltages with parabolic and linear functions, respectively, and calculate quantities corresponding to $B_X$ and $B_Y$ (Eq. (22)). Substituting $B_X$ and $B_Y$ into Eq. (23), we obtain $\Delta H_X$~3.03 Oe and $\Delta H_Y$~2.99 Oe, which match well with $a_J$ and $b_J$ used in the numerical calculations.

### B. In-plane magnetization systems

Numerical results of in-plane magnetized systems are shown in Figs. 3. The material parameters used are relevant for in-plane magnetized systems with a small perpendicular magnetic anisotropy.

Figure 3 shows results when an out of plane external field, slightly tilted ($\theta_H$=5 deg) from the film normal, is applied. The in-plane component of the tilted field is directed along the magnetic easy axis, which is the x-axis here ($\varphi_H$=0 deg, the in-plane anisotropy field ($H_A$) is ~–4 Oe). The open symbols represent the numerical results. The equilibrium magnetization angles ($\theta_0$, $\varphi_0$) are plotted against the slightly tilted out of plane field in Figs. 3(a) and 3(b), respectively. The magnetization direction reverses (Fig. 3(b)) due to the in-plane component of the tilted field. Figure 3(c) and 3(d) show the first and second harmonic Hall voltages whereas Fig. 3(e) and 3(f) display the first and second harmonic longitudinal voltages. The solid lines represent the analytical solutions, which agree well for the Hall voltages (Fig. 3(c, d)) but show a small deviation at low fields for the longitudinal voltages (Fig. 3(e, f)). The deviation is due to the non-linear (higher order) terms in $R_{XX}$ (Eq. (32)).

To reduce contributions from the non-linear terms in $R_{XX}$, we have used $a_J$=0.3 Oe and $b_J$=0.3 Oe to generate the effective field in Fig. 3 ($\hat{p} = (0,1,0)$ is assumed as before). For $a_J$=3 Oe and $b_J$=3 Oe, as used in the calculations shown in Fig. 2, the non-linear terms dominate the harmonic longitudinal voltages: the analytical solutions do not match the numerical calculations in this external field range (the solution shows better agreement at higher fields). Note that such effect is negligible for the harmonic Hall voltages (Figs. 3(c) and 3(d)) with $a_J$=3 Oe and $b_J$=3 Oe. The resulting components of the current induced effective field ($\Delta H_{X,Y,Z}$) are shown in the inset of Fig. 3(b). One can identify that $\Delta H_Z$ is the STT term, which changes its sign upon magnetization reversal, and $\Delta H_Y$ is the field-like term. Since the magnetization lies along the x-axis, $\Delta H_X$ is nearly zero.

We fit the numerically calculated harmonic Hall voltages vs. field with a linear function and use Eq. (30) to estimate $\Delta H_Y$ (see Fig. 3(d) inset for $V_{2\omega}^{-1}$ vs. the external field and the corresponding linear fit). We obtain $\Delta H_Y \sim 0.32$ Oe for both magnetization direction (pointing along +x and –x). This agrees well with $b_J$ used in the numerical calculations.

For the harmonic longitudinal voltages, we use Eq. (38) to obtain $\Delta H_Z$. Fitting the external field dependence of the first and second harmonic signals with parabolic and linear functions, we find $\Delta H_Z \sim 0.30$ Oe and $\sim -0.30$ Oe for magnetization pointing along +x and –x, respectively. Although these values match that of $a_J$, it should be noted that non-linear effects start to take place when $a_J$ and $b_J$ becomes large. To overcome this difficulty, one can apply a large out of plane field to force the magnetic moments to point along the film normal, and then simultaneously apply an in-plane field to evaluate the current induced torques using Eq. (23). Another option is to use a material/system that possesses large $H_A$, such as high aspect ratio nanowires, so that a transverse in-plane magnetic field can be applied to evaluate the STT term.

## V. Conclusion

We have derived analytical formulas that describe the adiabatic (low frequency) harmonic Hall and longitudinal voltages measurements when current induced spin orbit torques develop in magnetic heterostructures. We treat both out of plane and in-plane magnetized samples, taking into account the anomalous and planar Hall effects for the Hall voltage measurements and the anisotropic mangnetoresistance for the longitudinal voltage measurements. The derived forms are compared to numerical calculations using a macrospin model. The model used to describe out of plane samples agree well with the numerical calculations. For in-plane magnetized samples, although there are some difficulties in accurately evaluating the size of the current induced effective field, partly due to the non-linear effects that takes place when current is increased, we show that one can use the Hall and longitudinal voltages to evaluate the two orthogonal components of the effective field (spin transfer and field-like terms). Utilizing the harmonic voltage measurements can help gaining solid understanding of the spin orbit torques, which is key to the development of ultrathin magnetic heterostructures for advanced storage class memories and logic devices.

**Acknowledgements**

The author thanks Kevin Garello and Kyung-Jin Lee for fruitful discussions which stimulated this work, and Junyeon Kim and Seiji Mitani for helpful discussions. This work was partly supported by the Grant-in-Aid (25706017) from MEXT and the FIRST program from JSPS.

**References**

1. J. E. Hirsch, Phys. Rev. Lett. **83**, 1834 (1999).
2. S. F. Zhang, Phys. Rev. Lett. **85**, 393 (2000).
3. Y. A. Bychkov and E. I. Rashba, J. Phys. C **17**, 6039 (1984).
4. V. M. Edelstein, Solid State Commun. **73**, 233 (1990).
5. J. C. Slonczewski, J. Magn. Magn. Mater. **159**, L1 (1996).
6. L. Berger, Phys. Rev. B **54**, 9353 (1996).
7. S. Zhang and Z. Li, Phys. Rev. Lett. **93**, 127204 (2004).
8. A. Manchon and S. Zhang, Phys. Rev. B **78**, 212405 (2008).
9. T. Fujita, M. B. A. Jalil, S. G. Tan and S. Murakami, J. Appl. Phys. **110**, 121301 (2011).
10. K. W. Kim, S. M. Seo, J. Ryu, K. J. Lee and H. W. Lee, Phys. Rev. B **85**, 180404 (2012).
11. X. Wang and A. Manchon, Phys. Rev. Lett. **108**, 117201 (2012).
12. D. A. Pesin and A. H. MacDonald, Phys. Rev. B **86**, 014416 (2012).
13. E. van der Bijl and R. A. Duine, Phys. Rev. B **86**, 094406 (2012).
14. P. M. Haney, H. W. Lee, K. J. Lee, A. Manchon and M. D. Stiles, Phys. Rev. B **87**, 174411 (2013).
15. I. M. Miron, K. Garello, G. Gaudin, P. J. Zermatten, M. V. Costache, S. Auffret, S. Bandiera, B. Rodmacq, A. Schuhl and P. Gambardella, Nature **476**, 189 (2011).
16. L. Liu, C.-F. Pai, Y. Li, H. W. Tseng, D. C. Ralph and R. A. Buhrman, Science **336**, 555 (2012).
17. I. M. Miron, T. Moore, H. Szambolics, L. D. Buda-Prejbeanu, S. Auffret, B. Rodmacq, S. Pizzini, J. Vogel, M. Bonfim, A. Schuhl and G. Gaudin, Nat. Mater. **10**, 419 (2011).
18. K.-S. Ryu, L. Thomas, S.-H. Yang and S. Parkin, Nat. Nanotechnol. **8**, 527 (2013).
19. U. H. Pi, K. W. Kim, J. Y. Bae, S. C. Lee, Y. J. Cho, K. S. Kim and S. Seo, Appl. Phys. Lett. **97**, 162507 (2010).
20. I. M. Miron, G. Gaudin, S. Auffret, B. Rodmacq, A. Schuhl, S. Pizzini, J. Vogel and P. Gambardella, Nat. Mater. **9**, 230 (2010).
21. T. Suzuki, S. Fukami, N. Ishiwata, M. Yamanouchi, S. Ikeda, N. Kasai and H. Ohno, Appl. Phys. Lett. **98**, 142505 (2011).
22. J. Kim, J. Sinha, M. Hayashi, M. Yamanouchi, S. Fukami, T. Suzuki, S. Mitani and H. Ohno, Nat. Mater. **12**, 240 (2013).

23. K. Garello, I. Mihai Miron, C. Onur Avci, F. Freimuth, Y. Mokrousov, S. Blügel, S. Auffret, O. Boulle, G. Gaudin and P. Gambardella, arXiv:1301.3573 (2013).
24. S. Emori, U. Bauer, S.-M. Ahn, E. Martinez and G. S. D. Beach, Nat Mater **12**, 611 (2013).
25. X. Fan, J. Wu, Y. Chen, M. J. Jerry, H. Zhang and J. Q. Xiao, Nat Commun **4**, 1799 (2013).
26. P. Balaz, J. Barnas and J. P. Ansermet, J. Appl. Phys. **113**, 193905 (2013).
27. S. Zhang, P. M. Levy and A. Fert, Phys. Rev. Lett. **88**, 236601 (2002).
28. J. Z. Sun, Phys. Rev. B **62**, 570 (2000).
29. M. D. Stiles and J. Miltat, in *Spin Dynamics in Confined Magnetic Structures Iii* (Springer-Verlag Berlin, Berlin, 2006), Vol. 101, pp. 225.

**Figure Captions**

Fig. 1. Schematic illustration of the experimental setup. A Hall bar is patterned from a magnetic heterostructure consisting of a non-magnetic metal layer (gray), a ferromagnetic metal layer (blue) and an insulating oxide layer (red). The large gray square is the substrate with an insulating oxide surface. Definitions of the coordinate systems are illustrated together. $\vec{M}$ denotes the magnetization and $\vec{H}$ represents the external field.

Fig. 2. (a) Magnetization angle with respect to the film normal ($\theta_0$), (b) first harmonic Hall voltage and (d,f) second harmonic Hall voltage plotted against in-plane external field ($\theta_H$=90 deg). The field is directed along the x-axis ($\varphi_H$=0 deg) for (a,b,d) and along the y-axis ($\varphi_H$=90 deg) for (f). Open symbols show numerical calculations using the macrospin model. Solid/dashed lines represent the analytical solutions: (a) Eq. (17), (b) Eq. (21a), (d,f) Eq. (21b). (c,e) x, y, z component of the effective field used for the numerical calculations. Left (right) panel indicates the effective field when the magnetization is pointing along +z (-z). Parameters used in the numerical calculations: $H_K$=3160 Oe, $H_A$=-6 Oe, $\alpha$=0.01, $\gamma$=17.6 MHz/Oe, $a_J$=3 Oe, $b_J$=3 Oe, $\hat{p} = (0,1,0)$, $\Delta R_A$=1 Ω, $\Delta R_P$=0.1 Ω, $\Delta I$=1 A.

Fig. 3. (a) Polar ($\theta_0$) and (b) azimuthal ($\varphi_0$) angles of the magnetization, (c) first and (d) second harmonic Hall voltages ($V_{XY}$), (e) first and (f) second harmonic longitudinal voltages ($V_{XX}$) as a function of a slightly tilted out of plane field ($\theta_H$=5 deg). The in-plane component of the tilted field is directed along the x-axis ($\varphi_H$=0 deg). Open symbols show numerical calculations using

the macrospin model. Solid lines represent the analytical solutions: (a,b) Eq. (24), (c) Eq. (26), (d) Eq. (27), (e,f) Eq. (37). Inset of (b) shows the x, y, z component of the effective field used for the numerical calculations. Parameters used in the numerical calculations: $H_K$=-4657 Oe, $H_A$=-4 Oe, $\alpha$=0.01, $\gamma$=17.6 MHz/Oe, $a_J$=0.3 Oe, $b_J$=0.3 Oe, $\hat{p}=(0,1,0)$, $\Delta R_A$=1 $\Omega$, $\Delta R_P$=0.1 $\Omega$, $\Delta R_{MR}$=1 $\Omega$, $R_0$=0 $\Omega$, $\Delta I$=1 A.

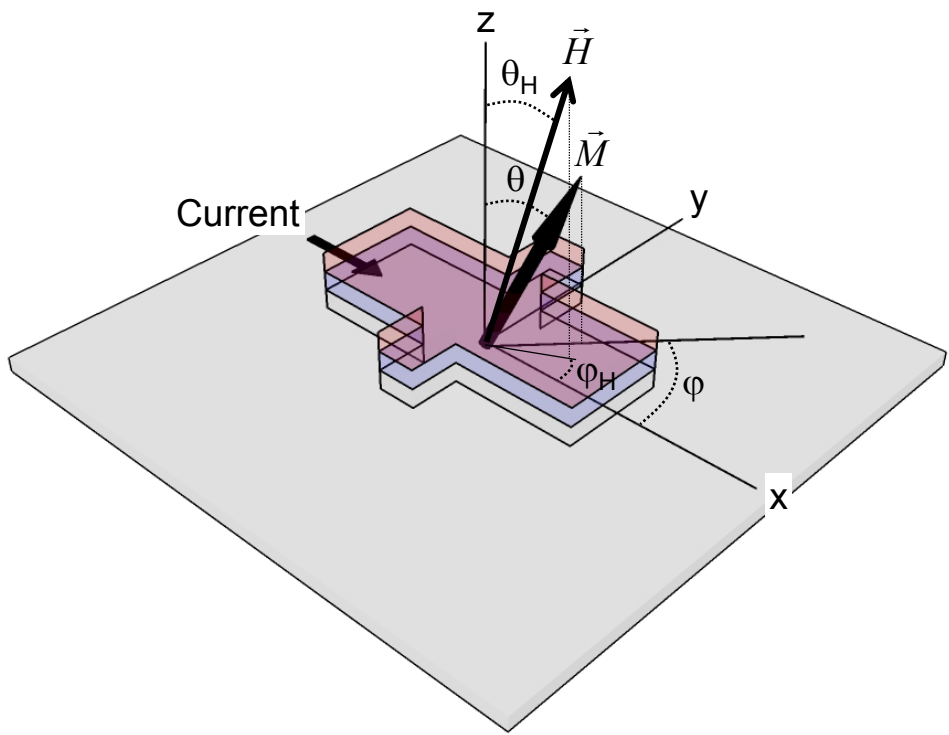

**Fig. 1**

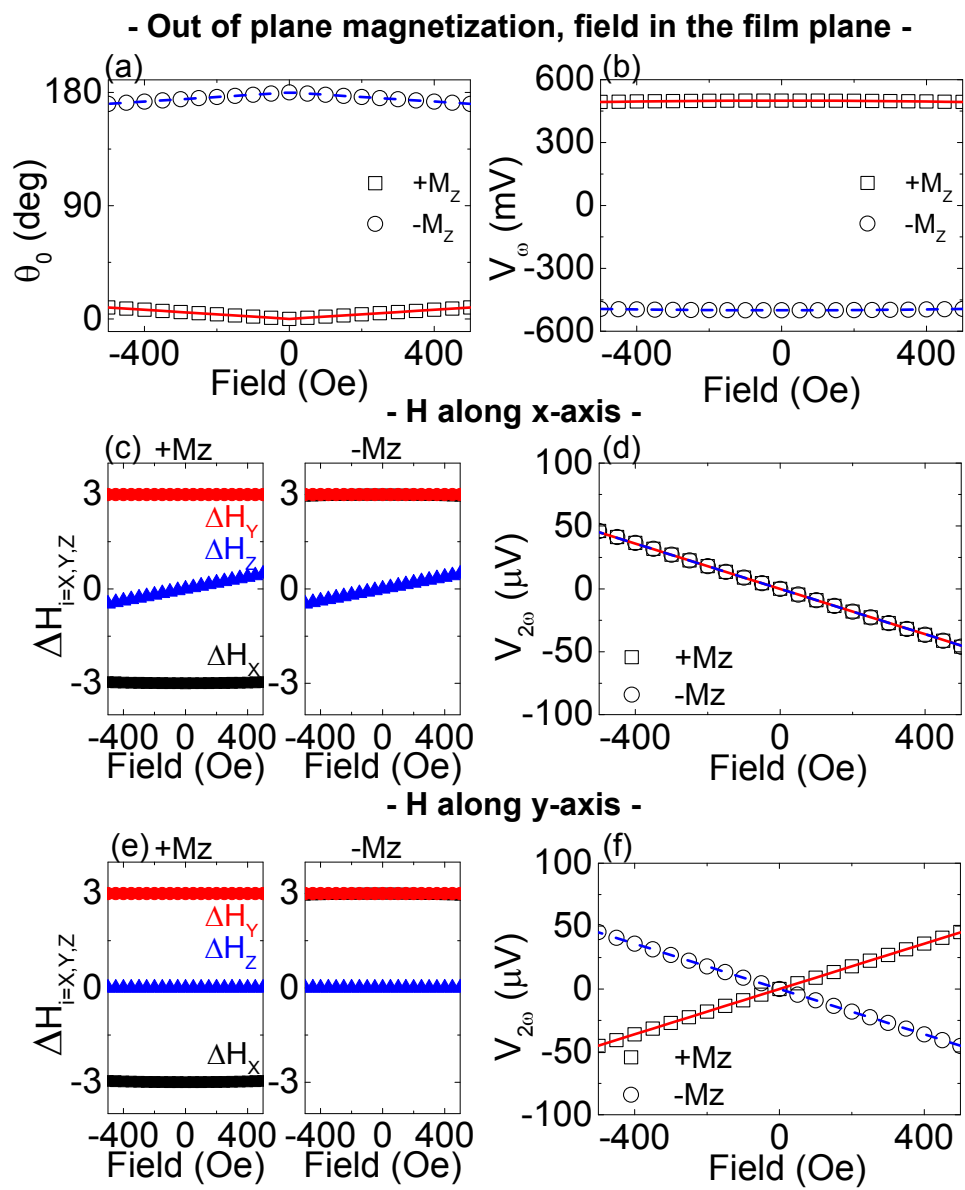

**Fig. 2**

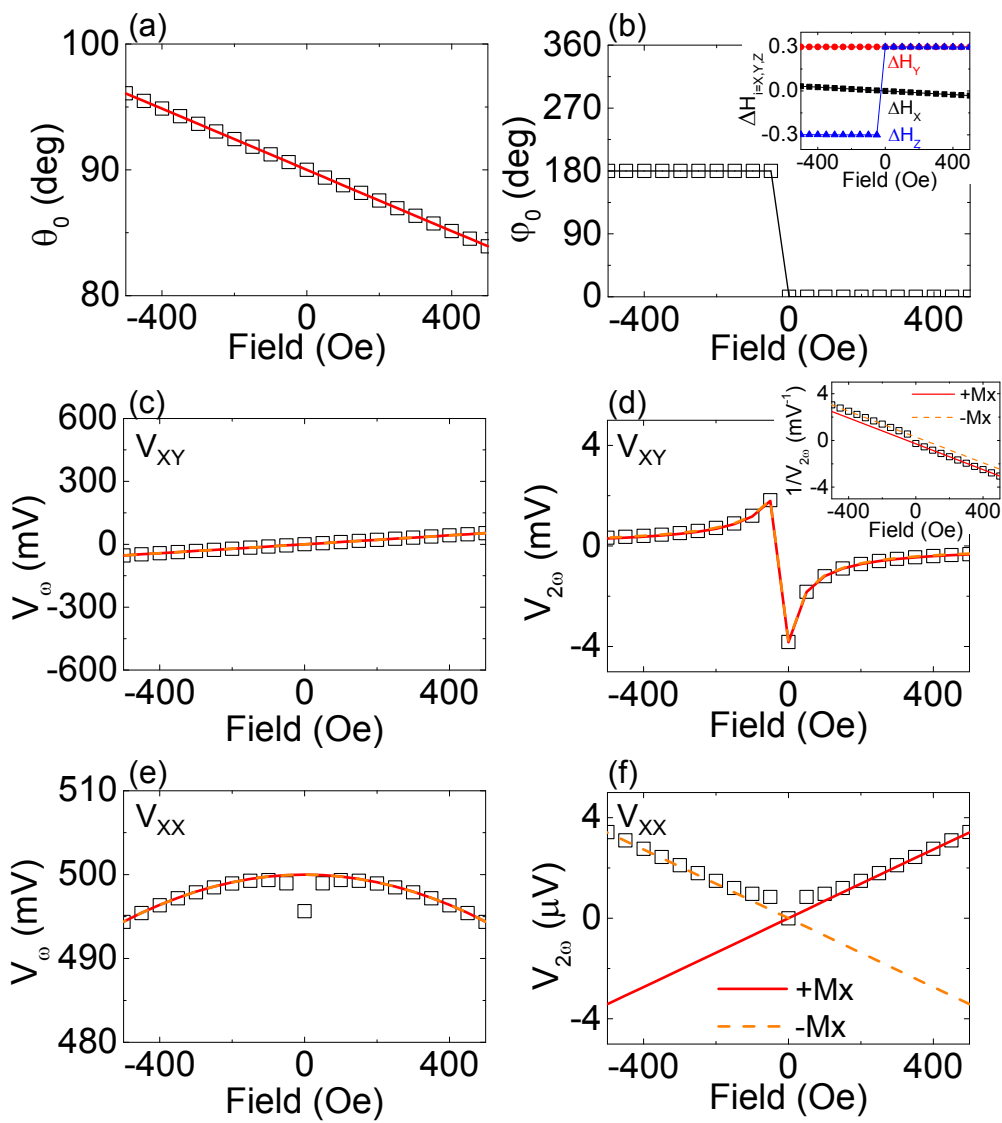

Fig. 3